\documentclass[pre,twocolumn]{revtex4}
\usepackage{graphicx}
\begin{document}
\title{Relaxation of nonspherical sessile drops towards equilibrium}
\author{Vadim S. Nikolayev}\email[email:]{vnikolayev@cea.fr}
\author{Daniel A. Beysens}
\affiliation{ESEME, Service des Basses Temp\'eratures, CEA-Grenoble (France)} \altaffiliation[Mailing
address:] {CEA-ESEME, Institut de Chimie de la Mati\`ere Condens\'{e}e de Bordeaux, 87, Avenue du Dr.
Schweitzer, 33608 Pessac Cedex, France}

\date\today

\begin{abstract}
We present a theoretical study related to a recent experiment on the coalescence of sessile drops. The study
deals with the kinetics of relaxation towards equilibrium, under the action of surface tension, of a
spheroidal drop on a flat surface. For such a non-spherical drop under partial wetting conditions, the
dynamic contact angle varies along the contact line. We propose a new non-local approach to the wetting
dynamics, where the contact line velocity depends on the geometry of the whole drop. We compare our results
to those of the conventional approach in which the contact line velocity depends only on the local value of
the dynamic contact angle. The influence on drop dynamics of the pinning of the contact line by surface
defects is also discussed.
\end{abstract}\maketitle

\section{Introduction}

At first glance, the motion of the gas-liquid interface along the solid surface is a purely hydrodynamic
problem. However, it attracted significant attention from the physicists since the work \cite{Huh}, which
showed an unphysical divergence that appears in the hydrodynamic treatment if a motion of a wedge-shaped
liquid slides along the solid surface. The reason for this divergence lies in the no-slip condition (i.e. zero
liquid velocity) at the solid surface. Being so common in hydrodynamics, this boundary condition is
questionable in the vicinity of the contact line along which the gas-liquid interface joins the solid. In the
absence of mass transfer between the gas and the liquid, the no-slip condition requires zero velocity for the
contact line that is supposed to be formed of the liquid molecules in the contact with the solid. It means
that, for example, an oil drop cannot move along the glass because of the no-slip condition! Of course, this
contradicts the observations.

The experiment \cite{Duss} demonstrated that the velocity on the liquid-gas interface is directed towards the
contact line during the contact line advance. The authors interpreted this result by the rolling
("caterpillar") motion of the drop \cite{footnote}. However, later theoretical study \cite{Mahad} shows that
such a motion is compatible with the no-slip condition on the non-deformable solid surface only for the
contact angles close to $180^\circ$.

The justification of the no-slip condition is well known \cite{DeGennes}: it is the excess of the attractive
force between the solid and the liquid molecules over the force between two liquid molecules. This
attraction has a tendency to prevent the motion of the liquid molecules adjacent to the solid. Obviously,
the same forces resist when these molecules are forced to move. In other words, some relatively large (with
respect to viscous dissipation) energy should be spent for this forcing.

Numerous microscopic theories (see e.g. \cite{DeGennes,CoalP,Sh,Sepp,YP,Cox,Voi,Blake}) propose different
phenomena as to be responsible for the contact line motion. However, no general theory has been agreed upon.
This situation is partly due to the scarceness of the information that can be extracted from the
experiments. Most of them deal with either drops with cylindrical symmetry (circular contact lines) or the
climbing of the contact line over a solid immersed into a liquid (straight contact line). In these
experiments, the contact line velocity $v_n$ measured in the normal direction does not vary along the
contact line on a macroscopic scale larger than the size of the surface defects. The experiments with
non-spherical drops where such a variation exists can give additional information. This information can be
used to test microscopic models of the contact line motion. To our knowledge, there are only two kinds of
investigated situations that feature the non-spherical drops. The first one is the sliding of the drop
along an inclined surface \cite{sliding}. The second concerns the relaxation of the sessile drops of
complicated shape towards the equilibrium shape of spherical cap. This latter case was studied
experimentally in \cite{CoalP} for water drops on the silanized silicon wafers at room temperature. The
present article deals with this second case.

The principal results of \cite{CoalP} can be summarized as follows:

(i) The relaxation of the drop from the elongated shape towards the spherical shape is exponential. The
characteristic relaxation time $\tau$ is proportional to the drop size. The drop size can be characterized by
the contact line radius $R^\ast$ at equilibrium when the drop eventually relaxes towards a spherical cap.

(ii)  The dependence of $\tau$ on the equilibrium contact angle $\theta$ is not monotonous:
$\tau(30^\circ)<\tau(53^\circ)$ and $\tau(53^\circ)>\tau(70^\circ)$.

(iii)  The relaxation is extremely slow. The capillary number $Ca=R^\ast\eta/(\tau\sigma)$ is of the order of
  $10^{-7}$, where $\sigma$ is the surface tension and $\eta$ is the shear viscosity.

Since the motion is not externally forced, a small $Ca$ shows that the energy dissipated in the vicinity of
the contact line is much larger than in the bulk of the drop.

The contact line motion is characterized by the normal component $v_n$ of its velocity. Many existing theories
result in the following relationship between $v_n$ and the dynamic contact angle $\theta$:
\begin{equation}\label{vnl}
  v_n=v_c\,F(\theta,\theta_s),
\end{equation}
where $\theta_s$ is the static contact angle, $v_c$ is a constant characteristic velocity and $F$ is a
function of two arguments, the form of which depends on the model used. For all existing models, the following
relation is satisfied
\begin{equation}\label{asym}
 F(\theta,\theta_s)=-F(\theta_s,\theta),
\end{equation}
which implies the trivial condition $F(\theta_s,\theta_s)=0$. It means simply that the line is immobile when
$\theta=\theta_s$.

The theories of Voinov \cite{Voi} and Cox \cite{Cox} correspond to
\begin{equation}\label{cox}
  F=\theta^3-\theta_s^3.
\end{equation}
There are many theories (see e.g. \cite{DeGennes}, \cite{Sh}) which result in
\begin{equation}\label{Fcos}
 F=\cos\theta_s-\cos\theta.
\end{equation}
In a recent model by Pomeau \cite{CoalP,YP}, it is proposed that
\begin{equation}\label{yp}
  F=\theta-\theta_s
\end{equation}
with the coefficient $v_c$ that depends on the direction of motion (advancing or receding) but not on the
amplitude of $v_n$.

Since the drop evolution is extremely slow, the drop shape can be calculated using the quasi-static argument
according to which at each moment the drop surface can be calculated from the constant curvature condition
and the known position of the contact line. The major problem is how to find this position. Independently
of the particular contact line motion mechanism, at least two approaches are possible. The first of them is
the ``local" approach \cite{CoalP}, which consists in the determination of the position of a given point of
the contact line from Eq.~(\ref{vnl}) where $\theta$ is assumed to be the \emph{local} value of the dynamic
contact angle at this point. Another, non-local approach is suggested in sec. \ref{nonloc}. Certainly, both
of this approaches should give the same result when $v_n$ does not vary along the contact line. However, we
show that the result is different in the opposite case.

The influence of surface defects on the contact line dynamics is considered in sec. \ref{Def}.

\section{Nonlocal approach to the contact line dynamics}\label{nonloc}

In this section we generalize another approach, suggested in \cite{DeCon},  for an arbitrary drop shape. This
approach postulates neither Eq.~(\ref{vnl}) nor a particular line motion mechanism. It simply assumes that
the energy dissipated during the contact line motion is proportional to its length and does not depend on the
direction of motion (advancing or receding). Then, at low contact line velocity, the leading contribution to
the energy dissipated per unit time (i.e. the dissipation function) can be written in the form
\begin{equation}
T=\oint{\xi \, v_n^2\over 2}\;{\rm d}l,\label{diss}
\end{equation}
where the integration is performed over the contact line and $\xi$ is the constant dissipation coefficient.
According to the earlier discussed experimental results \cite{CoalP}, the dissipation in the bulk is assumed
to be much smaller than that in the vicinity of the contact line.

Since we assume that most dissipation takes place in the region of the drop adjacent to the contact line,
our discussion is limited to the case where the prewetting film (that is observed for zero or very low
contact angles) is absent. This situation corresponds to the conditions of the experiment \cite{CoalP} where
the dropwise (as opposed to filmwise) condensation shows the absence of the prewetting liquid film. The
above assumption also limits the description to the partial wetting case. This assumption is also justified
by the experimental conditions under which it is extremely difficult to obtain macroscopic convex drops for
the contact angles $<30^\circ$ because of the contact line pinning \cite{CoalP}.  The main reason is that
the potential energy $U$ of the drop from Eq.~(\ref{U2}) goes to zero as the contact angle goes to zero. At
small contact angles $U$ is not large enough to overcome the pinning forces that originate from the surface
defects, see sec. \ref{Def}. Therefore, the macroscopic convex drops under consideration can not be observed
at small contact angles.

Generally speaking, the behavior of the drop obeys the Lagrange equation \cite{Land}
\begin{equation}
{{\rm d}\over{\rm d}t} \left({\partial {\cal L}\over\partial\dot{q_j}}\right)-{\partial {\cal L}\over\partial
q_j }=-{\partial T\over\partial\dot{q_j}},  \label{Lagr}
\end{equation}
where the Lagrangian ${\cal L}=K-U$ is the function of the generalized coordinates $q_j$ and of their time
derivatives which are denoted by a dot. The current time is denoted by $t$, $K$ is the kinetic energy, and
$U=U(q_j)$ is the potential energy. Since there is no externally forced liquid motion in this problem and the
drop shape change is extremely slow, we can neglect the kinetic energy by putting ${\cal L}=-U$. Then
Eq.~(\ref{Lagr}) reduces to
\begin{equation}
{\partial U\over\partial q_j }=-{\partial T\over\partial\dot{q_j}},  \label{UT}
\end{equation}
the expression applied first to the contact line motion in \cite{DeGennes}.

The potential energy of a sessile drop is \cite{DeCon}
\begin{equation}
U=\sigma(A_{VL}- A_{SL} \cos\theta_{eq})\label{UU},
\end{equation}
where $\sigma$ is the liquid surface tension, $A_{VL}$ and $A_{LS}$ are the areas of the vapor-liquid and
liquid-solid interfaces respectively and $\theta_{eq}$ is the equilibrium value of the contact angle. We
neglect the contribution due to the van der Waals forces because we consider macroscopic drops and large
contact angles $\ge 30^\circ$. For such drops the van der Waals forces influence the interface shape only in
the very close vicinity of the contact line and this influence can be neglected.

In general the static contact angle $\theta_s$ is not equal to $\theta_{eq}$ because of the presence of the
defects, a problem which will be treated in the section \ref{Def}. Meanwhile, we assume that
$\theta_s=\theta_{eq}$. The gravitational contribution is neglected in Eq.~(\ref{UU}) because the drops under
consideration are supposed to be small, with the radius much smaller than the capillary length. The volume of
a sessile drop is fixed. Its calculation provides us with another equation, which closes the problem provided
that the shape of the drop surface is known. The drop shape is determined from the condition of the
quasi-equilibrium which results in the constant curvature of the drop surface.

Usually, the wetting dynamics is observed either for the spreading of droplets with the shape of the spherical
cap, or for the motion of the liquid meniscus in a cylindrical capillary, or for the extraction of a solid
plate from the liquid \cite{DeGennes}. In all these cases, the contact line velocity $v_n$ does not vary along
the contact line and the dissipation function in the form Eq.~(\ref{diss}) results in the expression
\cite{DeCon}
\begin{equation}
v_n = {\sigma\over\xi}\,(\cos\theta_s-\cos\theta),  \label{cos}
\end{equation}
which is equivalent to Eq.~(\ref{vnl}), with the function $F$ taking the usual form~(\ref{Fcos}). One might
think that this equivalence confirms the universal nature of this expression. In the next section we show
that it is not exactly so because the non-local approach results in a different expression when $v_n$ varies
along the contact line.

\begin{figure}[htb]
  \begin{center}
  \includegraphics[height=6cm]{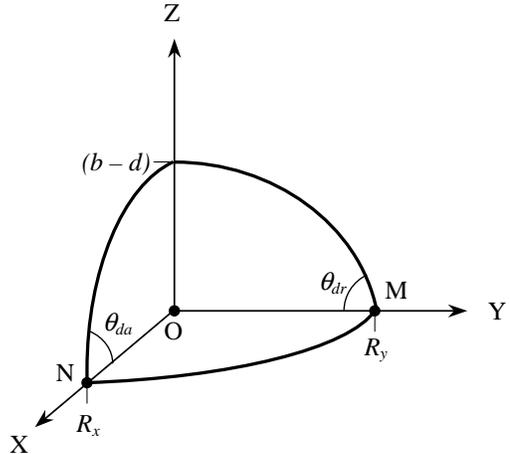}
  \end{center}
\caption{ Reference system to describe the 3D spheroidal cap. Only one quarter of it is shown. The surface is
described by Eq.~\ref{Ell}. The local contact angles at the points M and N are shown too.} \label{Quarter}
\end{figure}

Let us now apply the algorithm described above to the problem of drop relaxation. A shape for a non-spherical
drop surface of constant curvature can be found only numerically. In order to treat the problem analytically,
we approximate the drop shape by a spheroidal cap that is described in Cartesian $(x,y,z)$ coordinate system
by the equation
\begin{equation}
{\frac{x^2}{a^2}}+{\frac{y^2+(z+d)^2}{b^2}}=1  \label{Ell}
\end{equation}
at $z>0$, the plane XOY corresponding to the solid surface. The symmetry of the problem allows only quarter
of the drop (see Fig.~\ref{Quarter}) to be considered.
Since one of the parameters $(a,b,d)$ is fixed by the condition of the conservation of the drop volume that
can be calculated as
\begin{equation} V={\frac{\pi}{3}}{\frac{a}{b}}\,(2\,b^3-3\,b^2d+d^3),  \label{VV}
\end{equation}
there are only two free parameters left.

The time-dependent parameters $a$ and $b$ can be taken as generalized coordinates. However, it is more
convenient to use another set of parameters, $R_x$ and $R_y$, which are the half-axes of the ellipse that
form the base of the drop (see Fig.~\ref{Quarter}), $R_y>R_x$. They are related to $a$ and $b$ by the
equations
\begin{equation}
R_y^2+d^2=b^2,  \quad \mbox{and} \quad R_xb=R_ya,  \label{rel}
\end{equation}
that follow from Eq.~(\ref{Ell}). At the end of the relaxation
\begin{equation}
R_{x}=R_{y}=R\sin \theta_s\equiv R^\ast,
\end{equation}
where $R$ is the final radius of curvature of the drop. Therefore, during
the late stage
\begin{equation}
\begin{array}{l}
R_{x}=R^\ast\,(1-r_{x}) \\
R_{y}=R^\ast\,(1+r_{y})
\end{array}\label{RR}
\end{equation}
with $|r_{x,y}|\ll 1$. Some points of the contact line advance, some points recede. The dynamic contact angle
changes its value along the contact line. In particular, the point N$(R_x,0,0)$ in Fig.~\ref{Quarter} advance
and M$(0,R_y,0)$ recede. These points are extreme and their velocities have the maximum absolute values,
positive for N and negative for M. The dynamic contact angles ($\theta_{da}$: dynamic advancing contact angle
in N and $\theta_{dr}$: dynamic receding contact angle in M) also have the extreme values there. They can be
found from the equations
\begin{equation}
\begin{array}{l}
\cos\theta_{dr}=d/b, \\
\tan\theta_{da}=R_y^2\,/(d\,R_x),
\end{array}\label{angles}
\end{equation}
that reduce for $r_{x},r_{y}\ll 1$ to
\begin{equation}
\begin{array}{r}
\cos\theta_{dr}=\cos\theta_s+ \sin^2\theta_s  (2+\cos\theta_s)\cdot\\(2\,r_y-r_x)/3,\\
\cos\theta_{da}=\cos\theta_s- \sin^2\theta_s\,  [(2+4\,\cos\theta_s)\,r_x-\\(4-\cos\theta_s)\,r_y]/3.
\end{array}\label{coslin}
\end{equation}

Eq. \ref{UT}, written for the generalized coordinates $r_x$ and $r_y$ together with the expression for the
dissipation function (see Appendix \ref{A2}), implies the set of equations
\begin{equation}\left\{
\begin{array}{c}
3\,\dot{r}_x-\dot{r}_y=\tau_0^{-1}(B\,r_y-A\,r_x), \\
3\,\dot{r}_y-\dot{r}_x=\tau_0^{-1}(B\,r_x-A\,r_y),
\end{array}\right.\label{kin}
\end{equation}
where $\tau_0=\sigma R^\ast/\xi$ and the coefficients $A$ and $B$ are given by Eq.~(\ref{AB}) in Appendix
\ref{A2}. The solutions of Eqs. \ref{kin} read
\begin{eqnarray}
r_{x}(t) =[(r_{x}^{(i)}-r_{y}^{(i)})\,\exp(-t/\tau_s)+ \nonumber\\(r_{x}^{(i)}+r_{y}^{(i)})\,\exp(-t/\tau_n)]/2, \label{rx}\\
r_{y}(t) =[(r_{y}^{(i)}-r_{x}^{(i)})\,\exp(-t/\tau_s)+
\nonumber\\(r_{x}^{(i)}+r_{y}^{(i)})\,\exp(-t/\tau_n)]/2, \label{ry}
\end{eqnarray}
where $r_{x}^{(i)}$ and $r_{y}^{(i)}$ are the initial ($t=0$) values for $r_{x}$ and $r_{y}$ respectively,
and the relaxation times
\begin{eqnarray}
\tau_s= \tau_0/[\sin^2\theta_s\,(2+\cos\theta_s)],\\
\tau_n= 45\,\tau_0\,(1+\cos\theta_s)/[(108+41\,\cos\theta_s+\nonumber\\14\,\cos^2\theta_s
+17\,\cos^3\theta_s)(1-\cos\theta_s)].\label{taun}
\end{eqnarray}
The variables $r_{x}$ and $r_{y}$ are defined in Eq.~(\ref{RR}) in such a way that when
$r_{x}^{(i)}=-r_{y}^{(i)}$ the drop surface remains spherical during its relaxation. One can see from
Eqs.~(\ref{rx},\ref{ry}) that the evolution is defined entirely by the characteristic time $\tau _s$
(``spherical") in this case. When $r_{x}^{(i)}=r_{y}^{(i)}$, only $\tau _n$ (``non-spherical") defines the
drop evolution. In the real experimental situation where
$(r_{x}^{(i)}-r_{y}^{(i)})\ll(r_{x}^{(i)}+r_{y}^{(i)})$, the relaxation time $\tau_n$ alone defines the
relaxation of the drop as it follows from Eqs.~(\ref{rx}, \ref{ry}). Therefore $\tau_n$ should be associated
with the experimentally observed relaxation time.

\begin{figure}[htb]
  \begin{center}
  \includegraphics[height=6cm]{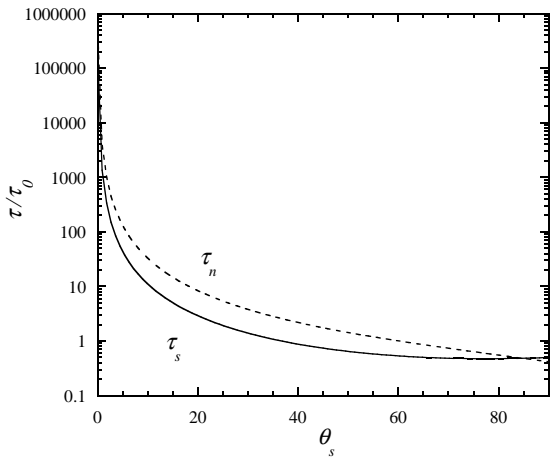}
  \end{center}
\caption{ The relaxation times $\tau_{s,n}$ versus the static contact angle $\theta_s$.}\label{times}
\end{figure}

The functions $\tau_{s,n}(\theta_s)$ are plotted in Fig.~\ref{times} assuming that $\xi$ is independent of
$\theta_s$.
Clearly, both $\tau _s$ and $\tau _n$ increase monotonically with $\theta_s$ in agreement with the observed
tendency for large contact angles.

It is interesting to check whether or not by applying the local approach of Eq.~(\ref{cos}) we recover the
non-local result for $v_n$. This is easy to do for the case $r_{x}^{(i)}=r_{y}^{(i)}$, i.e. when
$r_{x}=r_{y}$. In this case, the non-local model (\ref{kin}) implies $\dot{r}_x=-r_x/\tau_n$, and the contact
line velocities at the points M and N are
\begin{eqnarray}
v_n=&-{\sigma\over\xi}{\tau_0\over\tau_n}\,r_x  \quad\mbox{at the point M,}\label{vM}\\
v_n=&{\sigma\over\xi}{\tau_0\over\tau_n}\,r_x  \quad\mbox{at the point N.}\label{vN}
\end{eqnarray}
Since Eq.~(\ref{coslin}) results in
\begin{eqnarray}
(\cos\theta_s-\cos\theta)=-\sin^2\theta_s\,(2+\nonumber\\\cos\theta_s)\,r_x/3  \;\mbox{at the point M,}\label{cM}\\
(\cos\theta_s-\cos\theta)=\sin^2\theta_s\, (5\,\cos\theta_s-\nonumber\\2)\,r_x/3  \;\mbox{at the point
N,}\label{cN}
\end{eqnarray}
the local approach (\ref{cos}) implies that
\begin{eqnarray}
v_n=&-{\sigma\over\xi}{1\over 3}\sin^2\theta_s\,(2+\cos\theta_s)\,r_x  \quad\mbox{at the point M,}\label{vMl}\\
v_n=&{\sigma\over\xi}{1\over 3}\sin^2\theta_s\,(5\,\cos\theta_s-2)\,r_x  \quad\mbox{at the point
N.}\label{vNl}
\end{eqnarray}
The comparison of Eqs.~(\ref{taun}-\ref{vN}) with Eqs.~(\ref{vMl},\ref{vNl}) show that the results of the
local and the non-local approaches are different. However, one can verify that the results are the same in the
limit of very small $\theta_s$. For finite contact angles, the non-local approach is not equivalent to the
local approach. The main difference can be summarized as follows. The $v_n$ value that is obtained with our
non-local approach can be presented in the form (\ref{vnl}) common for the local approach. However, while the
characteristic velocity $v_c$ from Eq.~(\ref{vnl}) is constant in the local approach, it is \textit{a function
of the position on the contact line} in the non-local approach. Indeed, the comparison of
Eqs.~(\ref{vM}-\ref{cN}) with Eqs.~(\ref{vnl}, \ref{Fcos}) shows that
\begin{eqnarray}
v_c=&3\,{\sigma\over\xi}{\tau_0\over\tau_n}/[\sin^2\theta_s\,(2+\cos\theta_s)]  \;\mbox{at the point M,}\label{vMc}\\
v_c=&3\,{\sigma\over\xi}{\tau_0\over\tau_n}/[\sin^2\theta_s\,(5\,\cos\theta_s-2)]  \;\mbox{at the point
N.}\label{vNc}
\end{eqnarray}

We do not expect our model to be a good description for the contact angles close to $90^\circ$. The reason is
the limitation of the spheroid model for the drop shape. The spheroidal shape necessarily fixes $\theta
_{dr}=90^\circ$ when $\theta _{da}=90^\circ$ independently of the contact line velocity, which is incorrect.
In addition, the spheroid model does not work at all for $\theta _s>90^\circ$. One needs to find the real
shape of the drop (which is defined by constant curvature condition) to overcome these difficulties.

In order to estimate the limiting value for $\theta_s$ for which the spheroid model works well we mention
that the dynamic advancing and receding contact angles defined by Eq.~(\ref{coslin}) must satisfy the
inequality $\theta _{dr}\leq\theta_s\leq\theta _{da}$. By putting $r_{x}^{(i)}=r_{y}^{(i)}$ in Eqs.~(\ref{rx},
\ref{ry}) one finds that this inequality is satisfied when $\theta_s<66^\circ$. The last inequality provides
us with the limit of the validity for the spheroidal model.

To conclude this section we note that our non-local approach to the dynamics of wetting is not equivalent to
the traditional local approach. Both approaches allow the relaxation time to be calculated for a given
contact angle provided that the contact angle dependence of the dissipation coefficient $\xi$ is known.
Additional experiments are needed to reveal which approach is the most suitable. Under the assumption that the
$\xi(\theta_s)$ dependence (if any) is weak, we find that the relaxation time decreases with the contact
angle.

This result explains the decrease of the relaxation time at large contact angles observed in \cite{CoalP}. We
think that the opposite tendency observed for the small contact angles is related to the influence of the
surface addressed in the next section.

\section{Influence of the surface defects on the relaxation time}\label{Def}

The motion of the contact line in the presence of defects has been frequently studied (see \cite{DeGennes}
for a review). However, little is understood at the moment because the problem is very sophisticated. Most of
its studies deal with the influence of the defects on the static contact line (see, e. g. \cite{Mezard}) when
they are responsible for the contact angle hysteresis. The latter was studied in \cite{Garoff} and
\cite{Collet} for the wedge geometry which assumes the external forcing of the contact line. When the contact
line moves under the action of a force $f$, it encounters pinning on the random potential created by surface
defects. Thus the motion shows the ``stick-slip" behavior. It is characteristic for a wide range of physical
systems where pinning takes place and is the basis of the theory of dynamical critical phenomena, in which
the average contact line velocity is
\begin{equation}
v_n=v_c(f/f_c-1)^\beta,  \label{v}
\end{equation}
where the exponent $\beta$ is universal and $f_c$ is the pinning threshold. This expression is often applied
(see \cite{Wong} and refs. therein) to the contact line motion in the systems, where the geometry of the
meniscus does not depend on the dragging force. However, the values of $\beta$ vary widely depending on the
experimental conditions and do not correspond to the theoretical predictions. The motion of the contact line
during the coalescence of drops is even more complicated because the geometry of the meniscus is constantly
changing. Therefore, application of the expression (\ref{v}) is even more questionable in this situation.

\begin{figure}[htb]
  \begin{center}
  \includegraphics[height=6cm]{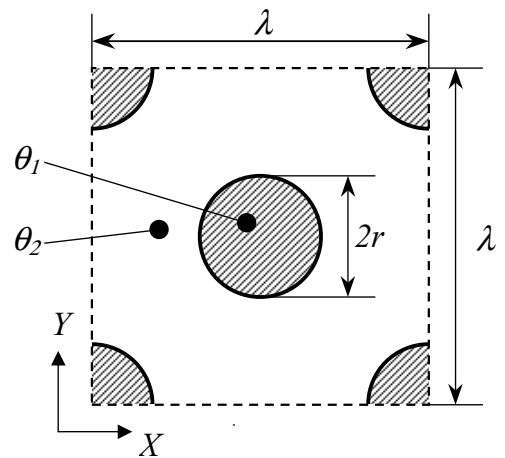}
  \end{center}
\caption{ Unit cell of the model defect pattern on the solid surface. The reference system and the values of
the contact angle inside the round spots ($\protect\theta_1$) and outside them ($\protect\theta_2$) are also
shown.} \label{Defects}
\end{figure}

In this section we employ the formalism developed in the previous section in order to understand the
influence of the surface defects on the relaxation time of the drop where the contact line is \textit{not
forced externally}. The surface defects are modeled by the spatial variation of the local density of the
surface energy, which can be related to the local value of the equilibrium contact angle
$\theta_{eq}(\vec{r})$ by the Young formula as was suggested in \cite{Garoff} to describe the static contact
angles. The expression (\ref{UU}) can be rewritten for this case in the form
\begin{equation}
U = \sigma A_{VL}-\sigma\int\limits_{(A_{LS})}\cos\theta_{eq}(\vec{r})\;{\rm d}\vec{r}.  \label{Ggen}
\end{equation}
The contribution of the defects and thus the deformation $\delta R_x$ of the contact line due to the defects
is assumed to be small. Then, in the first approximation that corresponds to the ``horizontal averaging"
approximation from \cite{Garoff}
\begin{equation}
U = U^{(0)}+\Delta U.  \label{U}
\end{equation}
The superscript $(0)$ means that the corresponding quantity is calculated for $\delta R_x=0$ and for the
constant value of the contact angle $\bar\theta_{eq}$ defined by the expression
\begin{equation}
\bar\theta_{eq} =
\arccos\left[{\frac{1}{A_{LS}^{(0)}}}\int\limits_{(A_{LS}^{(0)})}\cos\theta_{eq}(\vec{r})\;{\rm
d}\vec{r}\right]. \label{bt}
\end{equation}
Then
\begin{equation}
U^{(0)}=\sigma A_{VL}^{(0)}-\sigma A_{LS}^{(0)}\cos\bar\theta_{eq}.  \label{G0}
\end{equation}
It can be shown that the first-order correction to this value, which appears
due to the defects is
\begin{equation}
\Delta U= -\sigma\int\limits_{(A_{LS}^{(0)})}(\cos\theta_{eq}(\vec{r})-\cos\bar\theta_{eq})\;{\rm d}\vec{r}.
\label{DG}
\end{equation}
We accept the following model for defects because it is, on one hand, simple and, on the other hand, proven
\cite{Garoff} to be a good description for the advancing and the receding contact angles in the approximation
considered. The defects are supposed to be similar circular spots of radius $r$ arranged in a {\em regular}
spatially periodic pattern, $\lambda$ being the spatial period, the same in both directions, see
Fig.~\ref{Defects}.
The spots and the clean surface have the values of the equilibrium contact angle $\theta_1$ and
$\theta_2<\theta_1$ respectively. For this pattern, Eq. \ref{bt} yields
\begin{equation}
\bar\theta_{eq} = \arccos[\varepsilon^2\cos\theta_1+(1-\varepsilon^2)\cos\theta_2],  \label{cpp}
\end{equation}
the parameter $\varepsilon^2$ being the fraction of the surface covered by the defect spots. We consider the
case $r<\lambda/4$ in the following. Then it is obvious from Fig.~\ref{Defects}, that
\begin{equation}
\varepsilon^2=2\pi(r/\lambda)^2.  \label{eps}
\end{equation}

\begin{figure}[htb]
  \begin{center}
  \includegraphics[height=4cm]{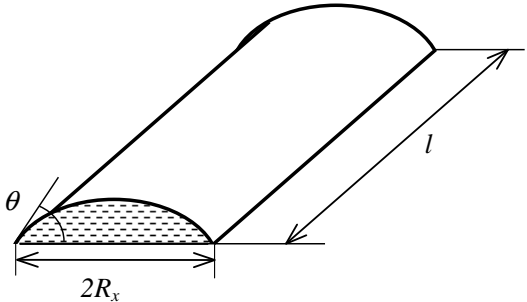}
  \end{center}
\caption{ Reference system to describe the 2D drop. The contact angle $\theta$ is shown too.} \label{2Ddrop}
\end{figure}

In the following, we will for simplicity treat a 2D sessile drop, i. e. a liquid stripe of infinite length,
the cross-section of which is the segment of a circle as shown in Fig.~\ref{2Ddrop}.
The volume $V$ of the stripe per its length $l$ does not change with time:
\begin{equation}
V/l= {R_x^2\over 2\,\sin^2\theta}\,(\theta-\sin\theta\,\cos\theta), \label{Vs}
\end{equation}
where $R_x$ is the half-width of the stripe, see Fig.~\ref{2Ddrop}. The dynamic contact angle $\theta$ can be
calculated from Eq.~(\ref{Vs}), provided that $R_x$ is known. It can be shown by the direct calculation of
$U^{(0)}$ (\ref{G0}) and the dissipation function $T$ (\ref{diss}) that Eq.~(\ref{UT}) with the substitution
$q_j\rightarrow R_x$ reduces to the equation
\begin{equation}
\dot{R_x}= {\sigma\over\xi}\,(\cos\bar\theta_{eq}-\cos\theta)- {1\over 2\,\xi l}\, {{\rm d}\Delta U\over{\rm
d}R_x} \label{dyn}
\end{equation}
The first-order correction to the drop energy $\Delta U$ can be calculated from Eq.~(\ref{DG}) by following
the guidelines of \cite{Garoff}. Its explicit expression for the chosen geometry is given in
Appendix~\ref{A3}.

\begin{figure}[htb]
  \begin{center}
  \includegraphics[height=6cm]{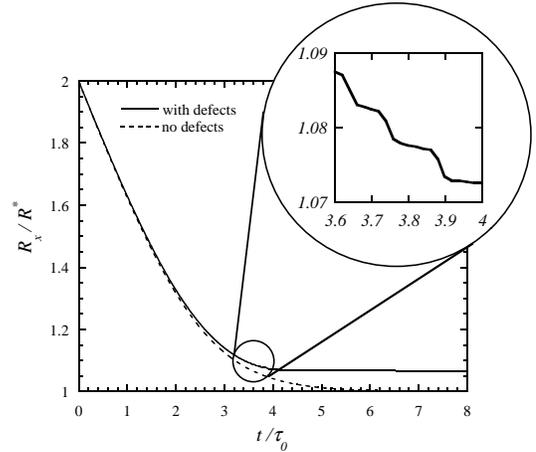}
  \end{center}
\caption{ Temporal evolution of the half-width $R_x$ of the drop with and with no defects with the same
initial ($t=0$) value of $R_x=2\,R^\ast$ and for $R^\ast=100\lambda$ and $\bar\theta\approx 55^\circ$. The
latter value corresponds to the defect radius $r=0.2\lambda$, $\theta_1=70^\circ$, and $\theta_2=50^\circ$.}
\label{Kinet}
\end{figure}

The kinetics of the relaxation is shown in Fig.~\ref{Kinet}. The relaxation kinetics for the drop on the
ideal substrate with the equilibrium value of the contact angle equal to the value of $\bar\theta_{eq}$ is
also shown for comparison. The half-width of the drop on the ideal substrate relaxes to its equilibrium value
$R^\ast$ that is related to the volume of the drop through Eq.~(\ref{Vs}) written for
$\theta=\bar\theta_{eq}$ and $R_x=R^\ast$. We chose $R^\ast/\lambda=100$ for Fig.~\ref{Kinet}.

The stick-slip motion is illustrated in the insert in Fig.~\ref{Kinet}.
Note that the contact line in its final position for the non-ideal case is pinned in a metastable state so
that the final 2D radius of the drop is larger than $R^\ast$. The final contact angle (the equilibrium
receding contact angle) thus differs from that for an ideal surface. Because the contact line is being stuck
on the defects, its motion is slowed down. However, the presence of defects does not change strongly the
relaxation time. It remains of the order of $\tau_0=R^\ast\xi/\sigma$ because this deceleration is
compensated by the acceleration during the slip motion.

Fig.~\ref{Kinet} shows the impact of the defects on the relaxation. The relaxation time appears to be
\emph{smaller} in the presence of defects than in the ideal case (no defects) because the contact line is
pinned by defects whereas it would have continued to move on the ideal surface.

It should be noted that this model is just a first step towards the description of contact line kinetics on a
non-ideal substrate. In reality, the different portions of the contact line slip at different moments in time
(cascades of slips are observed e. g. in \cite{Wong}). This means that the liquid flows in the direction
parallel to the contact line to the distances much larger than the defect size, i. e. the first-order
approximation is not adequate. The direction of this flow reverses frequently. This effect can lead to the
expression like Eq.~(\ref{v}) and to a large relaxation time.

\section{Conclusions}

This article deals with two important issues concerning contact line dynamics. First, it discusses the local
versus non-local approaches to contact line motion. While the local approach consists in postulating a direct
relationship between the normal contact line velocity and the dynamic contact angle \emph{at a given point of
the contact line}, the nonlocal approach starts from a more general hypothesis about the form of the
dissipation function of the droplet. These approaches give the same results for very small contact angles or
for the normal contact line velocity that does not vary along the contact line, which is the case of a drop
that has the shape of a spherical cap. In other cases (large contact angle, non-spherical drops) the results
of these two approaches differ. We carried out calculations assuming that the drop surface is a spheroid. In
reality, its surface is not a spheroid and has a constant curvature. More theoretical work is needed to
overcome this approximation.

The second issue treated in this article is the influence of surface defects on contact line dynamics. In the
approximation of a 2D drop, it is assumed that the contact line remains straight during its motion. In this
approximation, the stick-slip microscopic motion does not influence the average dynamics strongly. The
defects manifest themselves by changing the final position of the contact line by pinning it in a metastable
state. Therefore, the relaxation is more rapid than that on an ideally clean surface simply because it is
terminated earlier.

\appendix

\section{Derivation of the dynamic equations for $r_x$ and $r_y$}\label{A2}

We used the $Mathematica^{\rm TM}$ system for the analytical computations. We find first the dissipation
function $T$. The contact line can be described by the equation
\begin{equation}
F(x, y)=0\quad \mbox{with} \quad F(x, y)={\frac{x^2}{R_x^2}}+{\frac{y^2}{R_y^2}}-1, \label{cont}
\end{equation}
where $R_x$ and $R_y$ are time-dependent. By using the well-known formula of differential geometry
$v_n=-\dot{F}/|\nabla F|$, the integral (\ref{diss}) can be written in an explicit form. In order to obtain
the first order approximation for $T$, one can use the expansion (\ref{RR}). We need to keep only the
second-order terms. Since the integrand is a quadratic form with respect to $\dot{r_x},\dot{r}_y$, one can
put $r_{x},r_{y}=0$ in it. The resulting expression can be integrated to obtain the explicit expression for
the dissipation function
\begin{equation}
T={\xi\,\pi\,R^3\,\sin^3\theta_s\over 8}(3\,\dot{r}_x^2+3\,\dot{r}_y^2-2\,\dot{r}_x\,\dot{r}_y).\label{Tres}
\end{equation}
It is easy to find out that $T\ge 0$ always holds as it should be.

It is more difficult to obtain the drop interface area
\begin{equation}
A_{VL}=\int\limits_{A_{SL}}\sqrt{1+\left(\frac{\partial z}{\partial x}\right)^2+\left(\frac{\partial z}{\partial y}\right)^2}\;{\rm d}A  \label{AVL},
\end{equation}
where the function $z=z(x,y)$ is defined by Eq.~(\ref{Ell}). After the integration over $y$, Eq.~(\ref{AVL})
reduces to
\begin{equation}
A_{VL}=4\,b\int\limits_0^{R_x}\arctan\left(\frac{b}{d}\sqrt{1 - \frac{d^2}{b^2} - \frac{x^2}{a^2}}\right)\, \sqrt{1 - \frac{x^2}{a^2}\,\epsilon}\;{\rm d}x\label{AVL1},
\end{equation}
where $\varepsilon=1-R_y^2/R_x^2\sim(r_x-r_y)\ll 1$. The subsequent development of the integrand into the series over $\epsilon$ and its integration term-by-term results in
\begin{eqnarray}
A_{VL}= a\,\pi \,\left[ 2\,\left( b - d \right)-
    \frac{\epsilon}{6\,b^2}\left( 2\,b^3 - 3\,b^2\,d + d^3 \right) -\right.\nonumber\\
    \left.\frac{\epsilon^2}{160\,b^4}\left( 8\,b^5 - 15\,b^4\,d + 10\,b^2\,d^3 - 3\,d^5 \right) \,\right]\label{AVL2}.
\end{eqnarray}
This expression can be developed into a series with respect to $r_{x},r_{y}$ by using Eqs.~(\ref{VV},
\ref{rel}-\ref{RR}). Its substitution into Eq.~(\ref{UU}) leads to the explicit expression for $U$:
\begin{eqnarray}
U=\sigma\pi R^2 \Bigl\{2-3\,\cos\theta_s+\cos^3\theta_s+ \nonumber\\
{\sin^2\theta_s\over 8} \left[ A\left(r_x^2+r_y^2\right)  -
    2B\,r_x\,r_y \right]\Bigr\} \label{U2},
\end{eqnarray}
where
\begin{equation}
\begin{array}{c}
A=\left[(288+491\,\cos\theta_s+374\,\cos^2\theta_s+107\,\cos^3\theta_s)\cdot\right.\\\left.(1-\cos\theta_s)\right]/[45(1+\cos\theta_s)],\\
B=\left[(72+409\,\cos\theta_s+346\,\cos^2\theta_s+73\,\cos^3\theta_s)\cdot\right.\\\left.(1-\cos\theta_s)\right]/[45(1+\cos\theta_s)].
\end{array} \label{AB}
\end{equation}
It is easy to show that the expression in the square brackets in Eq.~(\ref{U2}) is positive for an arbitrary
$\theta_s$. It means that the function $U(r_x,\,r_y)$ has its minimum at the point $(r_x=0,\,r_y=0)$, i.e.
for the drop that has the shape of the spherical cap. This result was expected.

The substitution of Eqs.~(\ref{U2}, \ref{Tres}) into Eq.~(\ref{UT}) written for $q_j=(r_x,r_y)$ results in the
set of Eqs.~(\ref{kin}) and thus concludes their derivation.

\section{Expression for the first order correction to the drop energy caused by defects}\label{A3}

The accepted assumptions facilitate calculation of the $\Delta U(R_x)$. The resulting function is periodical
with the period $\lambda/2$, so that for $r<\lambda/4$ it can be presented in the form
\begin{equation}
{\frac{\Delta U}{2\,\sigma\,\rho\,\Delta c}}=-\left\{
\begin{array}{rl}
\varepsilon^2 \rho-\left[r^2\arcsin(\rho/r)+\right.\\\left.\rho(r^2-\rho^2)^{1/2}\right]/\lambda, &
0\leq \rho<r \\
\varepsilon^2\left(\rho-\lambda/4\right), & r\leq \rho< {\frac{\lambda}{2}}-r \\
\varepsilon^2\left(\rho-\lambda/2\right)+\\\Bigl\{r^2\arcsin\left[\left( \lambda/2-
\right.\right.\\\left.\left. \rho\right)/r\right]+ \left(\lambda/2-\rho\right)\Bigl[r^2-\\\left(\lambda/2
-\rho\right)^2 \Bigr]^{1/2}\Bigr\}/\lambda, & {\frac{\lambda}{2}}-r\leq \rho<{\frac{\lambda}{2}}
\end{array}
\right.  \label{DG1}
\end{equation}
where $\rho$ is the fractional part of $2R_x/\lambda$, multiplied by $\lambda/2$, and $\Delta
c=\cos\theta_2-\cos\theta_1$. Since $R_x\gg \lambda$, the presence of defects generates many local minima of
the function $U(R_x)$ near its global minimum. These minima represent the metastable states. According to this
model, the contact line is pinned in the minimum closest to its initial position.


\begin{thebibliography}{99}
\bibitem{footnote} The hydrodynamic flow calculations (see e.g. \cite{Huh,Sh,Sepp}), which use other
contact line motion models show that they are all compatible with these experiments.
\end{thebibliography}
\end{document}